\documentclass[%
aip,
amsmath,amssymb,
reprint,%
]{revtex4-1}

\usepackage{graphicx}
\usepackage{hyperref}
\hypersetup{
  colorlinks = true,
  urlcolor = blue,
  linkcolor = blue,
  citecolor = blue,
}
\usepackage{epstopdf}

\begin{document}

\title{Effect of annealing on the interfacial Dzyaloshinskii-Moriya interaction in Ta/CoFeB/MgO trilayers}

\author{R. A. Khan}
\email[Correspondence: ]{pyrak@leeds.ac.uk}
\affiliation{School of Physics and Astronomy, University of Leeds, Leeds LS2 9JT, United Kingdom}

\author{P. M. Shepley}
\affiliation{School of Physics and Astronomy, University of Leeds, Leeds LS2 9JT, United Kingdom}

\author{A. Hrabec}
\altaffiliation{Now at  Laboratoire de Physique des Solides, CNRS, Orsay, France}
\affiliation{School of Physics and Astronomy, University of Leeds, Leeds LS2 9JT, United Kingdom}

\author{A. W. J. Wells}
\affiliation{School of Physics and Astronomy, University of Leeds, Leeds LS2 9JT, United Kingdom}

\author{B. Ocker}
\affiliation{Singulus Technologies AG, 63796 Kahl am Main, Germany}

\author{C. H. Marrows}
\affiliation{School of Physics and Astronomy, University of Leeds, Leeds LS2 9JT, United Kingdom}

\author{T. A. Moore}
\affiliation{School of Physics and Astronomy, University of Leeds, Leeds LS2 9JT, United Kingdom}

\date{\today}

\begin{abstract}
The interfacial Dzyaloshinskii-Moriya interaction (DMI) has been shown
to stabilize homochiral N{\'e}el-type domain walls in thin films with perpendicular magnetic anisotropy and as a result permit
them to be propagated by a spin Hall torque. In this study, we demonstrate
that in Ta/Co$_{20}$Fe$_{60}$B$_{20}$/MgO the DMI may be influenced
by annealing. We find that the DMI peaks at $D=0.057\pm0.003$ mJ/m$^{2}$
at an annealing temperature of 230 $^{\circ}$C. DMI fields were
measured using a purely field-driven creep regime domain expansion
technique. The DMI field and the anisotropy field follow a similar trend
as a function of annealing temperature. We infer that the behavior
of the DMI and the anisotropy are related to interfacial crystal ordering
and B expulsion out of the CoFeB layer as the annealing temperature
is increased.

\end{abstract}

%\pacs{}

\maketitle

In thin magnetic multilayers current-driven domain wall (DW) motion
holds great potential for use in spintronic devices \cite{parkin2008magnetic,allwood2005magnetic,fukami2009low}. 
In multilayers with perpendicular magnetic anisotropy and structural inversion asymmetry
DW motion is governed by various torques generated by spin-orbit effects,
principally, the Rashba effect \cite{miron2010current,bychkov1984oscillatory,miron2011fast}
and the spin Hall effect \cite{emori2013current,liu2012current,dyakonov1971current}.
Furthermore, the presence of an antisymmetric exchange interaction, known
as the Dzyaloshinskii-Moriya interaction (DMI) \cite{dzyaloshinsky1958thermodynamic,moriya1960anisotropic},
is reported to influence the DW spin structure \cite{benitez2015magnetic} and thus its current-driven
dynamics \cite{thiaville2012dynamics,emori2013current,hrabec2014measuring,ryu2013chiral}.
Ta/CoFeB/MgO has been found to possess the DMI \cite{conte2015role,torrejon2014interface}.
Current flowing in the Ta layer can generate spin-orbit torque via
the spin Hall effect \cite{liu2012spin} that leads to magnetization
switching of the CoFeB and DW motion \cite{conte2014spin}.

Ta/CoFeB/MgO has a low density of DW pinning defects \cite{burrowes2013low},
and also forms part of a magnetic tunnel junction \cite{ikeda2010perpendicular}.
Thus, the prospect of efficient current-induced DW motion combined
with readout via tunnel magnetoresistance makes it promising for low
power memory or logic devices. Knowledge of the role of the DMI is
essential for understanding current-induced DW dynamics in this material.
The DMI originates at the heavy metal/ferromagnet interface where adjacent 
spins align through the exchange interaction mediated by a heavy atom with 
a large spin-orbit coupling. It manifests as an effective in-plane field, the DMI
field, acting locally on a Bloch wall, which is magnetostatically
favored, and converting it to a chiral N{\'e}el wall. The chirality arises
since the DMI field points in a specific direction as expressed by
$-\mathbf{D}\cdot(\mathbf{S}_{1}\times \mathbf{S}_{2})$, where $\mathbf{S}_{1}$ and $\mathbf{S}_{2}$
are neighboring spins and $\mathbf{D}$ is the DM vector.

Here we report how the DMI is affected by thermal annealing in Ta/CoFeB/MgO thin films, since annealing is generally
required to produce a strong perpendicular anisotropy in this system,
and sample heating is often used in nanofabrication procedures. 
The DMI was measured using a field-driven DW creep method
\cite{je2013asymmetric,hrabec2014measuring}. We report an optimum
annealing temperature for a maximum DMI in this material system. We then discuss the possible
underlying mechanisms with reference to the anisotropy field which was 
found to follow a similar behavior as a function of the annealing temperature.

\begin{figure}
\includegraphics[scale=0.29]{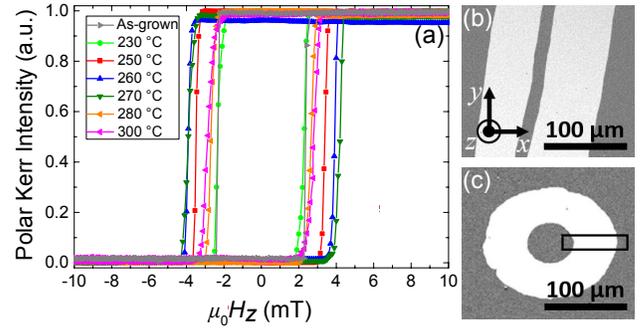}

\protect\caption{{\footnotesize{}(a) Polar MOKE hysteresis loops of Ta/CoFeB/MgO film
for different annealing temperatures. Kerr microscope difference images showing the propagation
of \textquotedblleft line\textquotedblright{} domains (b) and \textquotedblleft bubble\textquotedblright{}
domains (c) after a field pulse. The bright regions represent the
areas swept out by the DWs during the field pulse. The black rectangle in (c) marks an example
of a section inside which DW velocities are measured in each pixel
and averaged.\label{fig:Fig_1}}}

\end{figure}

The material system consists of Ta(5 nm)/Co$_{20}$Fe$_{60}$B$_{20}$(0.8
nm)/MgO(2 nm) deposited on a thermally oxidized Si wafer. The multilayer
was grown by sputtering using a Singulus TIMARIS/ROTARIS tool. A 5
nm capping layer of Ta was also deposited on top of the stack in order
to prevent degradation of the MgO layer in ambient conditions and
during annealing. The grown samples were then annealed at the desired
temperature (ramp rate of 5 $^{\circ}$C/min) for 2 hrs in vacuum
at a pressure of approximately $10^{-5}$ mbar. The as-deposited
and the annealed samples all exhibit a uniaxial magnetic anisotropy
perpendicular to the plane of the sample. This is shown by the square
magnetic hysteresis loops (FIG. \ref{fig:Fig_1}(a)) measured by polar magneto-optic
Kerr effect (MOKE) magnetometry. The as-deposited film shows \textquotedblleft line\textquotedblright{}
domains (FIG. \ref{fig:Fig_1}(b)), whereas, the same film when annealed at 200 $^{\circ}$C
exhibits \textquotedblleft bubble\textquotedblright{} domains (FIG. \ref{fig:Fig_1}(c)). Line domains occur
when an as-grown film is incompletely saturated, and reversal starts from two closely-spaced homochiral DWs \cite{benitez2015magnetic}.
In the present study, we measure the DMI only from annealed films where bubble domains are nucleated.
The domains were imaged using a wide field Kerr microscope
equipped with two electromagnets to generate an in-plane (IP) and an out-of-plane (OOP) magnetic
field simultaneously. Images captured before and after a field pulse
were subtracted. DW displacement is measured from the difference image
and the corresponding DW velocity is calculated by normalizing the
displacement by the pulse time. To reduce uncertainties, velocities
were calculated for each pixel of a section (black rectangle) and
averaged. This procedure was then repeated at least three times with
different pulse times and further averaged.

\begin{figure}
\includegraphics[scale=0.41]{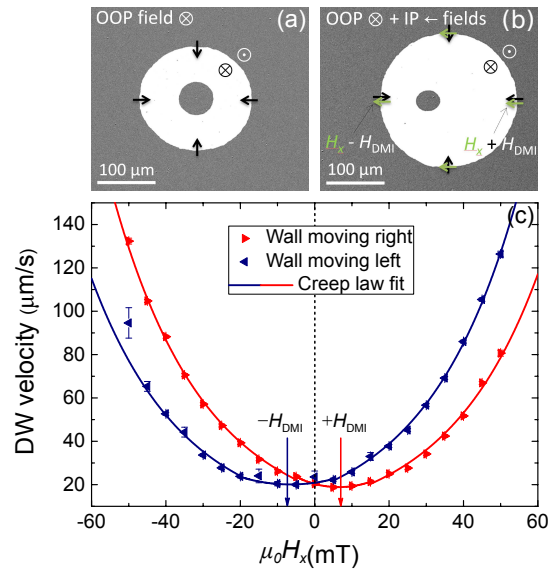}

\protect\caption{{\footnotesize{}Kerr microscope difference images showing: (a) isotropic expansion
of a nucleated bubble domain by an OOP field pulse $H_{z}$. The bright
region represents the area swept out by an up-down DW during the field
pulse. The black arrows represent the orientation of spins in the
center of an ideal N{\'e}el type DW due to the intrinsic DMI field $H_{\mathrm{DMI}}$;
(b) asymmetric expansion when the same method is performed but in
presence of a static IP field $H_{x}$. The green arrows represent the
action of the IP field on the DW spins. (c) DW velocity as a function
of IP field $H_{x}$ where the DW was driven with an OOP field pulse
$H_{z}$ of 1.4 mT. The error bars are from standard deviations of
the average values. The solid lines are fits of the creep scaling
law, Equation (\ref{eq:Creep Law}). The minimum points of the plots
mark the DMI fields $H_{\mathrm{DMI}}$ indicated by the arrows.\label{fig:Fig_2}}}
\end{figure}

The DMI field is measured using a field-driven DW creep method 
\cite{je2013asymmetric,hrabec2014measuring}. Using
field alone avoids the possibility of mixing with current-related
effects. In the case of a reverse nucleated circular bubble domain,
the DMI field $H_{\mathrm{DMI}}$ acts on the DW where it maintains radial
symmetry with respect to the axis of expansion, which is parallel
to the OOP field direction ($z$-axis in this case). Thus, the circular
domain expands in an isotropic way when an OOP field $H_{z}$ is applied,
as demonstrated by FIG. \ref{fig:Fig_2}(a). The symmetry is broken (FIG. \ref{fig:Fig_2}(b)) when the OOP
expansion is performed but in the presence of an applied IP field $H_{x}$.
This is because the applied IP field either adds to the DMI field
(right side in this case), or opposes it (left side). Thus, the effective
IP field acting on the DW is enhanced ($H_{x}+H_{\mathrm{DMI}}$) on one side,
causing an increase in the velocity, compared to the DW velocity on
the other side where the effective field is diminished ($H_{x}-H_{\mathrm{DMI}}$).

The effect of the DMI field on the DW velocity can be readily observed
by plotting the DW velocity as a function of IP field $H_{x}$, for
a fixed OOP field which drives the DW. FIG. \ref{fig:Fig_2}(c) shows the data for
a sample which has been annealed at 280 $^{\circ}$C. It can be seen
that the two DW velocity plots (red and blue for DWs moving right
and left, respectively) shift away from $H_{x}=0$ in opposite directions.
We take the offset of the minimum from $H_{x}=0$ as a measure of $H_{\mathrm{DMI}}$
since at this point $H_{x}$ exactly cancels out $H_{\mathrm{DMI}}$ resulting in
the lowest DW velocity. This velocity is non-zero since the OOP field
is still driving the DW. The minimum velocity, and therefore $H_{\mathrm{DMI}}$
can be easily identified from the plot without the need of any further
analysis. However, in order to identify $H_{\mathrm{DMI}}$ to a high 
degree of precision, the data points are fitted with the creep law of DW dynamics, 
which assumes that the DW is a 1D elastic interface moving in a 2D weakly disordered
medium and that its velocity increases exponentially as a function
of the driving force \cite{lemerle1998domain,chauve2000creep}. The
creep law is expressed as:

\begin{equation}
v=v_{0}\exp[-\zeta(\mu_{0}H_{z})^{-\mu}],\label{eq:Creep Law}
\end{equation}
where $\mu$ is the creep exponent which takes the value of $1/4$
for field-driven DW motion \cite{lemerle1998domain,kim2009interdimensional,metaxas2007creep},
the prefactor $v_{0}$ is the characteristic speed, and $\zeta$ is
a scaling factor and is expressed as:

\begin{equation}
\zeta=\zeta_{0}[\sigma(H_{x})/\sigma(0)]^{1/4},
\end{equation}
where $\zeta_{0}$ is a scaling constant and $\sigma$ is the DW energy
density. The constants $v_{0}$ and $\zeta_{0}$ were extracted from
the intercept and gradient of a linear fit of the plot of $\ln v$
vs $(H_{z})^{-1/4}$ at $H_{x}=0$.

The DW energy density $\sigma$ is a function of the applied IP field
$H_{x}$ \cite{je2013asymmetric}, and takes the form of

\begin{equation}
\sigma(H_{x})=\sigma_{0}-\frac{\pi^{2}\Delta\mu_{0}^{2}M_{\mathrm{s}}^{2}}{8K_{\mathrm{D}}}(H_{x}+H_{\mathrm{DMI}})^{2}
\end{equation}
for the condition $|H_{x}+H_{\mathrm{DMI}}|<4K_{\mathrm{D}}/\pi\mu_{0}M_{\mathrm{s}}$. This
is when the effective IP field acting on the DW $(H_{x}+H_{\mathrm{DMI}})$
is not sufficient to completely transform a Bloch wall into a N{\'e}el
wall, i.e. at relatively low applied IP fields. In this scenario the
spin structure of the DW is in a Bloch-N{\'e}el mixed state. Otherwise,
when a DW is fully transformed into a N{\'e}el wall, the DW energy density
is expressed as:

\begin{equation}
\sigma(H_{x})=\sigma_{0}+2K_{\mathrm{D}}\Delta-\pi\Delta\mu_{0}M_{\mathrm{s}}|H_{x}+H_{\mathrm{DMI}}|.
\end{equation}

In these expressions, $\sigma_{0}$ is the Bloch wall energy density
and is expressed as $\sigma_{0}=4\sqrt{AK_{0}}$, where $A$ is
the exchange stiffness, taken to be 10 pJ/m,  and $K_{0}=\mu_{0}H_{\mathrm{K}}M_{\mathrm{s}}/2$
is the effective anisotropy, where $\mu_{0}H_{\mathrm{K}}$ is the measured
effective anisotropy field (discussed later); $\Delta=\sqrt{A/K_{0}}$
is the DW width; $M_{\mathrm{s}}=(6.50\pm0.04)\times10^{5}$ A/m is the saturation
magnetization and is measured by a Quantum Design SQUID-VSM; $K_{\mathrm{D}}=N_{x}\mu_{0}M_{\mathrm{s}}^{2}/2$
is the magnetostatic shape anisotropy of the wall with $N_{x}$ as
the demagnetization prefactor \cite{tarasenko1998bloch}. FIG. \ref{fig:Fig_2}(c) 
shows that the experimental data fits well with the
DW creep model (solid lines). This model was fitted to the data for all the 
samples annealed at different temperatures. The DMI fields were 
consequently extracted from the fits. We then calculated the effective
DM constant $D$ by using the expression \cite{thiaville2012dynamics} $D=\mu_{0}H_{\mathrm{DMI}}M_{\mathrm{s}}\Delta$.

\begin{figure}
\includegraphics[scale=0.31]{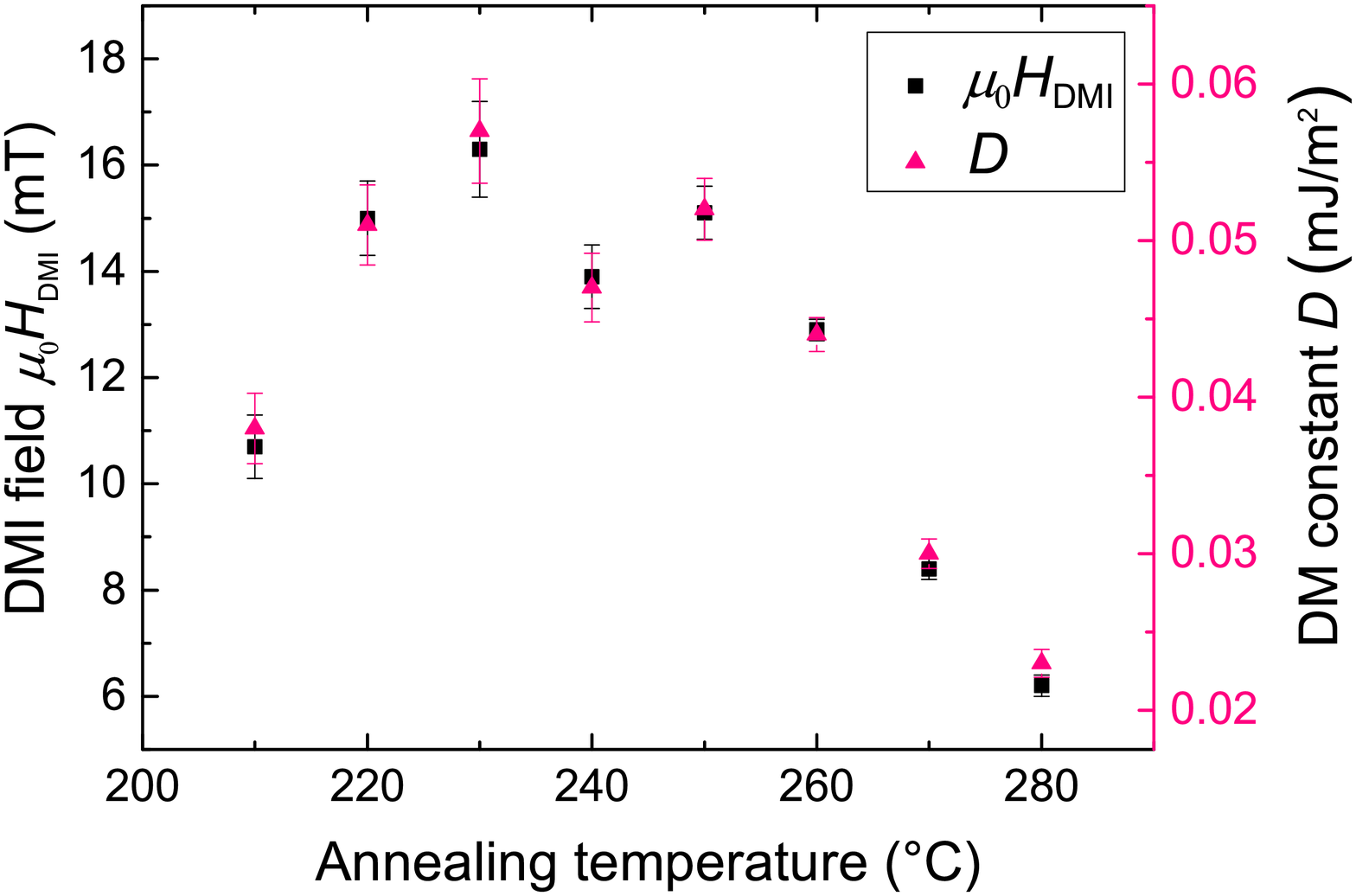}

\protect\caption{{\footnotesize{}Plot of the DMI field $\mu_{0}H_{\mathrm{DMI}}$ (black squares)
and the DM constant $D$ (pink triangles) as a function of annealing
temperature. The uncertainties in the DMI fields are from standard 
deviations of the averages. \label{fig:Fig_3}}}
\end{figure}

FIG. \ref{fig:Fig_3} shows how the magnitude of the DMI field $H_{\mathrm{DMI}}$ (black)
and subsequently the DM constant $D$ (pink) varies as a function of
annealing temperature. We find that the DMI gradually rises from
$D=0.038\pm0.002$ mJ/m$^{2}$ at a temperature of 210 $^{\circ}$C, 
reaches a peak value of $D=0.057\pm0.003$ mJ/m$^{2}$ at 
230 $^{\circ}$C, and then decreases as the temperature is increased
further.

\begin{figure}
\includegraphics[scale=0.76]{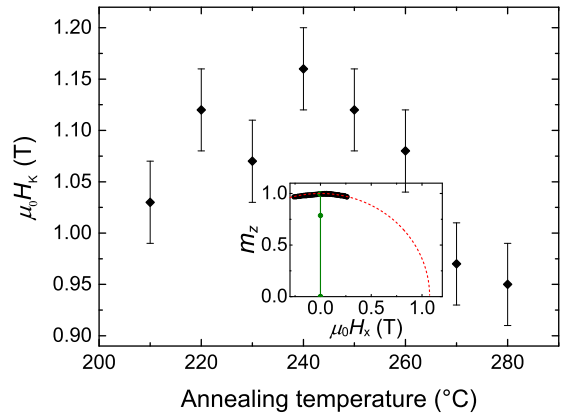}

\protect\caption{{\footnotesize{}Effective anisotropy field $\mu_{0}H_{\mathrm{K}}$ vs annealing
temperature. Inset: Plot of (normalized) OOP magnetization $m_{z}$
vs IP field $\mu_{0}H_{x}$. The OOP magnetization was probed using
a Kerr microscope in the polar settings. Before applying the IP field,
the magnetization was saturated along the easy $z$-axis using an OOP
field $\mu_{0}H_{z}$ (green data points). This ensured a maximum
value of $m_{z}$, while $m_{x}=0$. The red dashed line is the extrapolation
of the plot assuming that the magnetization rotates coherently as
the IP field is increased. Hence, the $x$-intercept represents the anisotropy
field.\label{fig:Fig_4}}}
\end{figure}

The anisotropy field $\mu_{0}H_{\mathrm{K}}$ is measured magneto-optically
for a low field range over which the magnetization rotates coherently.
In this method, as illustrated in FIG. \ref{fig:Fig_4} (inset), the Kerr microscope
is set up in the polar configuration so that the OOP component of
the magnetization $m_{z}$ is probed. In this configuration, $m_{z}$
is measured continuously while an IP field $H_{x}$ is applied to
rotate the magnetization from the easy (OOP) to the hard axis (IP) \cite{Shepley2015modification}.
At $H_{x}=0$, the magnetization is saturated in the $z$-direction
(easy axis) using an OOP field $H_{z}$ (green points) resulting in
the maximum value of $m_{z}$ (while $m_{x}=0$). Now as $H_{x}$
is increased, the magnetization starts to rotate towards the $x$-direction
(hard axis) and thus $m_{z}$ gets smaller in magnitude until nucleation
of domains starts to occur causing a sharp drop in $m_{z}$ (not shown).
Thus, the low field data is extrapolated to obtain the anisotropy
field from the $x$-intercept assuming that the magnetization rotates
coherently, i.e. $\mathbf{m}=m_{z}^{2}+m_{x}^{2}=1$, within this
low field regime. The uncertainty is obtained from the quality of
the fit and by performing repeated measurements.

We find that the anisotropy field follows a similar trend (FIG. \ref{fig:Fig_4}) as the DMI, 
with regard to the annealing temperature, peaking in magnitude
at about the same temperature as the DMI peak. Such a behavior of
the anisotropy field was also reported by Avci $et$ $al.$ \cite{avci2014fieldlike},
although for a smaller temperature range.
The initial rise in the anisotropy field is due to an increase in the crystal ordering of
the CoFeB and MgO layers due to annealing \cite{cardoso2005characterization,cui2013perpendicular}.
Crystallization of the CoFeB layer is also brought about by the diffusion
of B, due to annealing, out of the CoFeB and into the adjacent layers.
This was reported by Lo Conte $et$ $al.$ through chemical depth profiling
\cite{conte2015role}. An increased ordering of these two layers leads
to a rise in the anisotropies at the MgO/CoFeB and CoFeB/Ta interfaces,
both of which contribute to the PMA of the stack. However, further increasing
the annealing temperature causes a decrease in the magnetic
anisotropy. We attribute this to a combined effect of B deposition
\cite{conte2015role}, and intermixing \cite{cui2013perpendicular}
at both the MgO/CoFeB and CoFeB/Ta interfaces due to annealing at
relatively high temperatures.

Since the DMI and the anisotropy field follow a similar trend with
respect to the annealing temperature, we infer that similar mechanisms 
underpin both these phenomena. Since the DMI is sensitive to the atomic 
arrangements at the interface \cite{lavrijsen2015asymmetric,hrabec2014measuring}, an improved ordering of the atoms at 
the Ta/CoFeB interface brought about by annealing is the reason 
for the initial enhancement of the DMI. At higher annealing
temperatures the accumulation of B at the Ta/CoFeB interface becomes
significant and essentially weakens the interaction between the atoms
of the Ta and the CoFeB layers. Furthermore, annealing at higher temperatures
also leads to intermixing at the interface which has been reported
\cite{yang2015anatomy} to be detrimental for the DMI. Thus, these
two factors together contribute to the lowering of the strength of
the DMI at relatively high annealing temperatures.

The obtained magnitude and sign of $D$ agrees well with previous reports
\cite{conte2015role,torrejon2014interface} on Ta/CoFeB/MgO stacks.
The chirality of the DW can be deduced from the directions of the
OOP and IP fields. The DWs in this system are determined to have a
right-handed chirality and thus the sign of $D$ is positive.

In conclusion, we have demonstrated how the interfacial DMI in Ta(5
nm)/Co$_{20}$Fe$_{60}$B$_{20}$(0.8 nm)/MgO(2 nm) multilayer is
affected by annealing temperature. We measured DMI fields via the
field-driven expansion of magnetic domains and found that the DMI
peaks at $D=0.057\pm0.003$ mJ/m$^{2}$ at a temperature of 
230 $^{\circ}$C. This behavior is related to interfacial 
crystal ordering and segregation of B out of the CoFeB
layer and consequent accumulation at the Ta/CoFeB interface, as the
anisotropy field is found to follow a similar trend and peaks in
magnitude at around the same temperature as the DMI field.

This work has been funded by the European Community under the 
Marie-Curie Seventh Framework program - ITN \textquotedblleft WALL\textquotedblright  (Grant no. 608031). 
Equipment funding has been provided by U.K. EPSRC; Grant no. EP/K003127/1 
for the Kerr Microscope, and Grant no. EP/K00512X/1 for the SQUID-VSM. 
The authors would like to thank O. Cespedes and G. Burnell for helpful discussions. 

\bibliography{References}% Produces the bibliography via BibTeX.

\end{document}